\documentstyle[12pt]{article}
\def\rsim{>\kern-2.5ex\lower0.85ex\hbox{$\sim$}\ }

\catcode`@=11
\def\seceqa{\@addtoreset{equation}{section}
           \def\theequation{1.\arabic{equation}}}
\def\seceqb{\@addtoreset{equation}{section}
           \def\theequation{2.\arabic{equation}}}
\def\seceqc{\@addtoreset{equation}{section}
           \def\theequation{3.\arabic{equation}}}
\def\seceqd{\@addtoreset{equation}{section}
           \def\theequation{4.\arabic{equation}}}
\def\seceqe{\@addtoreset{equation}{section}
           \def\theequation{5.\arabic{equation}}}
\def\seceqaa{\@addtoreset{equation}{section}
           \def\theequation{A\arabic{equation}}}
\def\seceqab{\@addtoreset{equation}{section}
           \def\theequation{B\arabic{equation}}}
\catcode`@=12

\begin{document}

\title{Nuclear Binding Effects in Relativistic Coulomb Sum Rules}
\author{D.S.~Koltun and T.C.~Ferr\'ee \\
{\it Department of Physics and Astronomy}\\
{\it University of Rochester}\\
{\it Rochester, New York 14627-0171}}
\maketitle
\vskip 0.5 true in

\begin{abstract}
We extend the formulation of relativistic Coulomb sum rules to account
for the average effects of nuclear binding on the initial and final
states of ejected nucleons.  Relativistic interactions are included
by using a Dirac representation adapted from a vector-scalar field
theory.  The scalar field reduces the effective nucleon mass $M^*$ and
increases the relativistic effects of recoil and Fermi motion.
We consider two models for the off-shell behavior of the nuclear
electromagnetic current, and demonstrate that the sum rule is accurate
for applications to data over the interesting range of $M^*$ and
three momentum $q$.  We further indicate that the form of the sum
rule is sufficiently general to accommodate a broad class of off-shell
form factor models.\hfill\break
\hfill\break
\end{abstract}
\vskip 0.5 true in
\centerline{Submitted to {\em Physical Review C}}

\vfil
\eject

\section{Introduction}
\setcounter{equation}{0}
\seceqa

The subject of this paper is the Coulomb sum rule (CSR) for inelastic
electron scattering~\cite{mvvh}, in a reformulation which extends its
validity to relativistic momentum transfers $q\!\rsim\!M$, where
$q\!\equiv\!|{\bf q}|$ is the three-momentum transfer to the target
nucleus, and $M$ is the nucleon mass.\footnote{For a derivation of the
nonrelativistic CSR in second quantization, see Ref.~\cite{fw}.}
The purpose of this work is to provide
practical methods for the analysis of the longitudinal $(e,e^\prime)$
response of nuclear targets, at the higher energies now available
at the Continuous Electron Beam Accelerator Facility (CEBAF) and
other electron accelerators.  In a previous paper~\cite{fk}, we
derived a relativistic Coulomb sum rule (RCSR) which incorporates
the relativistic effects of nucleon recoil at large $q$, as
well as of Fermi motion.  In the present work we extend this
theoretical approach to include the effects of nuclear interactions
on the RCSR.

In our recent article, we discussed in detail the assumptions under
which a RCSR could be derived.  The basic approximation is that only
nucleon, i.e., as opposed to antinucleon, degrees of freedom enter the
Coulomb response in the spacelike regime accessible by $(e,e^\prime)$
experiments.  We call this the ``nucleons-only'' approximation; it
ignores effects of antinucleons, but includes fully the relativity of
the nucleons.  Following conventional treatments, we adopt an
impulse approximation which ignores explicit contributions from the
exchange of charged mesons.  We include nucleon anomalous moments and
elastic form factors, which are important at higher energies.  The final
assumption is less conventional, but absolutely necessary.  In order to
derive a {\em non-energy-weighted} sum rule, it must be possible to
factor {\em all} dependence on the photon energy $\omega$ from the current
matrix element, otherwise dynamical effects enter into the Coulomb
response function and complicate the isolation of correlation effects.
In this paper, we restrict our attention to form factor models which
have factorable dependence on $\omega$.

We define the nuclear Coulomb sum in terms of the Coulomb response
function $W_C(\omega,{\bf q})$ and the proton electric form factor
$G_{E,p}(Q^2)$:

        \begin{equation}\Sigma({\bf q})\equiv\int_{\omega_{el}^+}
        ^q d\omega\ {W_C(\omega,{\bf q})\over G_{E,p}^2(Q^2)},
        \label{aa}\end{equation}

\noindent where the lower limit $\omega_{\rm el}^+$ excludes the
quasielastic peak, and  the upper limit $q$ restricts the integration
to spacelike four-momenta.  Under the assumptions of Ref.~\cite{fk}
mentioned above, we obtain a RCSR which can be expressed in terms of
one- and two-body contributions:

        \begin{equation}\Sigma({\bf q})\equiv
        \Sigma^{(1)}({\bf q})+C({\bf q})
        +\Sigma^{(2)}_{\rm un}({\bf q})
        ,\label{ab}\end{equation}

\noindent where the one-body contribution is of the form

        \begin{equation}\Sigma^{(1)}({\bf q})
        \equiv2\sum_{{\bf p}\sigma}
        n_\sigma({\bf p})\ r_\sigma({\bf p},{\bf q}),
        \label{ac}\end{equation}

\noindent $C({\bf q})$ includes two-body correlation information in
momentum space, and $\Sigma_{\rm un}^{(2)}({\bf q})$ is the uncorrelated
two-body part which is related to the square of the nuclear elastic form
factor, as discussed in Ref.~\cite{fk}.

In the one-body term (\ref{ac}), $n_\sigma({\bf p})$ is the nucleon
momentum distribution function for isospin projection $\sigma$ and one
spin projection, and $r_\sigma({\bf p},{\bf q})$ is a kinematic factor
which arises due to relativistic nucleon recoil and Fermi motion in the
target.  In the nonrelativistic limit ($q\!<\!\!<\!M$), we have $r_p
\!\rightarrow\!1$ and $r_n\!\rightarrow\!0$, which leads to
$\Sigma^{(1)}({\bf q})\!\rightarrow\!Z$; then the sum rule (\ref{ab})
is the result of Ref.~\cite{mvvh}.  The relativistic effects
are all in the functions $r_\sigma({\bf p},{\bf q})$, representing
recoil of the struck nucleon and Fermi motion in the target ground
state.  In Ref.~\cite{fk} we further showed that (\ref{ac}) does
not depend strongly on the details of $n_\sigma({\bf p})$, but
only on the lowest momentum moments, e.g., $\langle{\bf p}^2
\rangle_\sigma$.  This leads to a method of evaluating (\ref{ac})
accurately in a weakly model-dependent manner, and in principle
permits the extraction of the correlation function $C({\bf q})$
from the the experimentally measured Coulomb sum (\ref{aa}),
using (\ref{ab}).

In the present paper, we investigate what changes are required in the sum
rule when the Coulomb sum $\Sigma({\bf q})$ is modified by relativistic
two-body interactions.  We study the effect of the average interaction in
the nucleus using the mean field approach of quantum hadrodynamics, a
relativistic field theory for nuclear physics~\cite{sw}.  In particular,
we consider QHD-I, which includes vector and scalar isoscalar mesons only.
We then reformulate the Coulomb sum rule of Ref.~\cite{fk} so that results
(\ref{aa})--(\ref{ac}) have a similar structure, but with modifications
reflecting the mean-field effects of these relativistic interactions.

These modifications have two main effects on the RCSR:  First, there
are kinematical effects resulting from mean-field interactions of the
nucleons, which are represented in QHD-I by a reduction in the effective
nucleon mass $M^*$ in the medium.  Consequences of a reduced effective
nucleon mass $M^*$ for the Coulomb sum have been considered previously
in a Fermi gas model~\cite{mat,ddb}.  These calculations include
interactions of both initial and final plane-wave nucleon states with
the mean fields, through the effective mass $M^*$, which can be
interpreted as binding in the initial state, and final
state interactions of the ejected nucleon with the nucleus.  Chinn,
Picklesimer and Van Orden~\cite{cpa,cpb} have studied the effects of
final state interactions on the Coulomb response of a Fermi gas, using
more realistic interactions, and have seen similar effects to those
seen due to $M^*/M\!<\!1$.   Second, the electromagnetic coupling of
the nucleon in medium may be modified by the mean fields, entering
through the off-shell behavior of the nucleon elastic form factors.
Both modifications introduce a degree of model-dependence in the
RCSR which is not present in the nonrelativistic formulation, nor
in the relativistic formulation of Ref.~\cite{fk}.  We show how these
features can be incorporated into the theory to allow the evaluation of
the one-body contribution $\Sigma^{(1)}({\bf q})$, and the subsequent
extraction of the two-body correlation function $C({\bf q})$ from
the measured Coulomb response.

This paper is organized as follows: In Section 2 we introduce the basic
formalism to include relativistic mean-field effects in the Coulomb
sum rule (RCSR).  We introduce two models (F and G)
for the electromagnetic charge operator, and investigate the resulting
behavior connected to different off-shell assumptions for the
nucleon elastic form factors.  In Section 3, we derive a modified version
of the RCSR (\ref{aa})--(\ref{ac}), concentrating on
the explicit changes to the one-body term $\Sigma^{(1)}({\bf q})$ in
off-shell Models F and G.  In Section 4, we illustrate the operation of
the sum rule in a simple nuclear system: uniform nuclear matter treated
in the mean-field approximation, with nuclear binding effects incorporated
using QHD-I.  We examine the sensitivity of the RCSR to $M^*$ and to
the choice of off-shell models (F and G), focussing on the convergence
of the moment expansion in each case.  We further demonstrate
that the particular form of the RCSR given here is applicable to {\em both}
models, and argue that the same form should also be valid for a broad
class of form factor models.  In Section 5, we draw conclusions, give
guidelines for the application of the RCSR to data, and indicate important
directions for future work.


\section{Formalism}
\setcounter{equation}{0}
\seceqb

In this section, we review the formalism for electron scattering from
nuclei, as it pertains to our development of relativistic Coulomb sum
rules.  We first give some standard results of the plane-wave
impulse approximation (PWIA), which is used for the analysis of
$(e,e^\prime)$ experiments on nuclei, in a single-particle basis which
accounts for the vector and scalar interactions of QHD-I.  We then give
two models for the nucleon form factors off shell, which will be used
in the next section to illustrate the sensitivity of the sum rule to
different off-shell assumptions.

\subsection{PWIA in $M^*$ basis}

We begin with the differential cross section for the scattering of
ultrarelativistic electrons from nuclear targets, which is commonly
written in the form

        \begin{equation}{d^2\sigma\over d\Omega^\prime dE^\prime}=
        {d\sigma_{\!_M}\over d\Omega^\prime}\
        \biggl[{Q^4\over{\bf q}^4}W_C(\omega,{\bf q})+\biggl({1\over2}
        {Q^2\over{\bf q}^2}+{\rm tan}^2{\theta\over2}\Bigr)
        W_T(\omega,{\bf q})\biggr],\label{ba}\end{equation}

\noindent where $q^\mu\!=\!(\omega,{\bf q})$ is the four-momentum
transferred from the electron to the nucleus via virtual photon
exchange, and $Q^2\!\equiv\!|{\bf q}|^2\!-\!\omega^2\!>\!0$.
The longitudinal contribution in (\ref{ba}) has been expressed in
terms of the Coulomb response function

        \begin{equation}W_C(\omega,{\bf q})\equiv\sum_f
        |\langle f|{\hat J}_0(q)|i\rangle|^2
        \ \delta(\omega-E_f+E_i),\label{bb}\end{equation}

\noindent where $|i\rangle$ and $|f\rangle$ denote initial
\footnote{For notational simplicity, we assume a nondegenerate target
ground state; the results are easily generalized to unpolarized targets
with $J\!\not=\!0$.} and final nuclear many-body states, respectively.

In general, the electromagnetic current density operator ${\hat J}_\mu
(q)$ may include contributions from both nucleons and charged mesons.
Meson exchange current (MEC) contributions have been considered by
Schiavilla et.~al.~\cite{sch}, for example, but are not included here.
Including only nucleons with electromagnetic form factors, the current
density operator can be written in the form

        \begin{equation}{\hat J}_\mu(q)\equiv\int d^3x\
        e^{i{\bf q\cdot x}}\ {\bar{\hat\psi}}({\bf x})
        \Gamma_\mu(q){\hat\psi}({\bf x}),\label{bc}\end{equation}

\noindent where ${\hat\psi}({\bf x})$ is the (Schr\"odinger picture)
field operator for a point Dirac nucleon, and $\Gamma_\mu$ represents
the electromagnetic coupling at the $\gamma NN$ vertex.  There is also
a sum over proton and neutron isospin projections, which will be
suppressed until needed.  Any effects of charged mesons not included
in the current operator (\ref {bc}) may be interpreted as two-body
effects, as discussed in Ref.~\cite{fk}.

To proceed to a relativistic sum rule, we shall expand the field operators
${\hat\psi}({\bf x})$ of (\ref{bc}) in a plane wave basis, i.e., in a PWIA.
In Ref.~\cite{fk}, we took this to be the free basis, i.e., the momentum
eigenstates of the free Dirac equation.  In the present paper, we shall
instead expand the field operators in a basis of plane waves moving in the
presence of uniform (isoscalar) Lorentz scalar and vector potentials, such
as would be generated by scalar and vector mesons in a relativistic field
theory of nuclei, e.g., QHD-I.  With this modification, the theoretical
development is formally similar to that of Ref.~\cite{fk}.  The consequences
of the change of basis show up after the nucleons-only approximation, in
which antinucleon degrees of freedom are removed approximately from the
spacelike Coulomb response.  In the following, nucleons-only will
refer to nucleons in the presence of scalar and vector potentials in the
target.  The effect of this modification is to include efficiently
the effects of these mean-field potentials in both the initial and final
target states.

We therefore use the plane-wave solutions of the Dirac equation

        \begin{equation}\Bigl[\gamma^\mu(i\partial_\mu
        -g_v V_\mu)-(M-g_s\phi)\Bigr]
        \psi(x)=0,\label{bd}\end{equation}

\noindent where the scalar and vector potentials are written in the forms
$g_s \phi$ and $g_v V_\mu$, respectively, corresponding to the scalar and
vector fields $\phi$ and $V_\mu$, and the associated coupling constants.
Here, as in the mean-field solution for uniform nuclear matter at rest,
$V_\mu\!=\!\delta_\mu^0 V_0$, and $V_0$ and $\phi$ are constants.  The
solutions to (\ref{bd}) are discussed in detail in Ref.~\cite{sw}.
In the case of uniform fields, the energy eigenvalues of (\ref{bd})
take the simple form $E_{\bf p}^{(\pm)}\!=\!g_v V_0\pm E_{\bf p}^*$,
where $E_{\bf p}^*\!\equiv\!\sqrt{{\bf p}^2\!+\!{M^*}^2}$ and the
effective nucleon mass is defined $M^*\!\equiv\!M\!-\!g_s\phi$.
With the nucleons-only approximation, we need only consider
positive-energy solutions to (\ref{bd}).  These are plane wave
solutions of momentum ${\bf p}$, which obey the equation

        \begin{equation}\Bigl[\gamma^0 E_{\bf p}^*-{\bf \gamma}\cdot
        {\bf p}-M^*\Bigr]u_s({\bf p})=0.\label{be}\end{equation}

\noindent
Explicit forms for the interacting solutions $u_s({\bf p})$ can be
obtained directly from the free solutions, given in Appendix A of
Ref.~\cite{fk}, by making the replacement $M\!\rightarrow\!M^*$.
Since the vector potential $g_v V_0$ appears additively in the nucleon
eigenenergy, it does not appear in (\ref{be}) or its plane-wave spinor
solutions.

In Ref.~\cite{fk} we argued that the Coulomb response function (\ref{bb}) for
 spacelike ($\omega\!<\!|{\bf q}|$) photon exchange is dominated by nucleon
($NN$) contributions to the current matrix elements, and that antinucleon
(${\bar N}{\bar N}$) and pair ($N{\bar N}$) terms could be neglected.
This is an exact result for a uniform free Fermi gas, and leads to the
nucleons-only approximation for interacting nuclear systems.  The
presence of a strong scalar field in the nucleus induces mixing of free
$N$ and ${\bar N}$ states, particularly for nucleons of high momentum,
as in the final states of (\ref{bb}).  However, transforming to the
plane-wave basis formed from the solutions of (\ref{be}), hereafter
referred to as the ``$M^*$-basis'', removes this mixing by the potentials.
The nucleons-only approximation is again adopted for interacting nuclei,
but here refers to nucleons of mass $M^*$.  This use of the $M^*$-basis
for final states implies substantial interaction of excited (ejected)
nucleons before leaving the target, and is probably a better assumption
for large nuclei than for small.  Then the electromagnetic current
operator takes the form\footnote{For simplicity, we discretize the
sum over the momentum {\bf p}.}

        \begin{equation}{\hat J}_\mu(q)\simeq\sum_{\bf p}\sum_{ss^\prime}
        {{\bar u}_{s^\prime}({\bf p\!+\!q})\over\sqrt{2E^*_{\bf p+q}}}
        \Gamma_\mu(q){u_{s}({\bf p})\over\sqrt{2E^*_{\bf p}}}
        a^\dagger_{{\bf p+q} s^\prime}a_{{\bf p} s},
        \label{bf}\end{equation}

\noindent where $a^\dagger_{{\bf p}s}$ and is a creation operator for
a nucleon with spin projection $s$ and momentum ${\bf p}$.
The energy denominators in (\ref{bf}) reflect the normalization of the
plane-wave spinors to $2E_{\bf p}^*$ particles/volume.  The formal
derivation of the RCSR now follows closely that of Ref.~\cite{fk}, once
we have discussed the form of the electromagnetic vertex operator
$\Gamma_\mu(q)$.

\subsection{Off-shell nucleon form factors}

It is conventional to express the $\gamma NN$ vertex operator in the form

	\begin{equation}\Gamma_\mu(q)=F_1\gamma_\mu+i{\kappa\over2M}
	F_2\sigma_{\mu\nu}q^\nu,\label{bg}\end{equation}

\noindent where $\kappa$ is the nucleon anomalous magnetic moment, $M$ is
the free nucleon mass, and $F_1$ and $F_2$ are the Dirac and anomalous
form factors, respectively.  In general, $F_1$ and $F_2$ are scalar
functions of $p$, $p^\prime$ and $q$.  This form is sufficiently general
for matrix elements between nucleon states (antinucleons excluded) as
in the nucleons-only PWIA.  Since the tensor $\sigma_{\mu\nu}$ is
antisymmetric, only the three-momentum {\bf q} enters explicitly the
Coulomb operator $\Gamma_0$.  Any remaining dependence the photon
energy $\omega$ enters $\Gamma_0$ {\it only} through the form
factors $F_1$ and $F_2$.

For scattering from a free nucleon, the form factors $F_1$ and $F_2$
depend only on the scalar $Q^2$.  To ensure the correct charge and
magnetic moments for free nucleons, they are normalized at $Q^2\!=\!0$
according to $F_{1p}(0)\!=\!F_{2p}(0)\!=\!F_{2n}(0)\!=\!1$ and
$F_{1n}(0)\!=\!0$.  The form factors $F_1$ and $F_2$ are obtained from
$(e,e^\prime)$ scattering data, usually in terms of the more convenient
Sachs electric and magnetic form factors $G_E$ and $G_M$:

	\begin{eqnarray}F_1(Q^2)
	&=&{G_E(Q^2)+\tau G_M(Q^2)\over1+\tau}\nonumber\\
	&&\label{bh}\\ \kappa F_2(Q^2)
	&=&{G_M(Q^2)-G_E(Q^2)\over1+\tau},
	\nonumber\end{eqnarray}

\noindent where $\tau\!\equiv\!Q^2/4M^2$, and $Q^2\!\equiv\!
-(k_\mu-k_\mu^\prime)^2$, i.e., the photon four-momentum is determined
by the four-momentum transfer at the $\gamma ee$ vertex.
For a single nucleon in free space, both energy and momentum are
conserved at the $\gamma NN$ vertex as well, and we have
$p^\prime_\mu\!=\!p_\mu\!+\!q_\mu$.  See Appendix A for an alternative
form of (2.7) which is expressed in terms of the Sachs form factors.
We use the standard parameterization of the Sachs form factors,
along with the assumption $G_{En}(Q^2)\!=\!0$, as discussed in
Ref.~\cite{fk}.  This choice has the convenient feature that all
Sachs form factors are proportional, which satisfies a condition
assumed in our derivation of a {\em non-energy-weighted} sum rule
in Ref.~\cite{fk}.

For interacting nucleons in a nucleus, one needs information about the
form factors off their free mass shell, which is simply not known.
This leaves considerable freedom to extrapolate off shell from the known
on-shell forms.  Given no other information, a common approach has been
to assume that the dependence on $Q^2$ of the form factors $F_1(Q^2)$
and $F_2(Q^2)$ does not change off shell from that for a free nucleon.
This implies, at least in part, a separation of the dynamics of the
electromagnetic structure of the nucleon from that of the nucleus.
However, even this assumption does not uniquely specify the off-shell
form factors, because of the freedom to transform $\Gamma_\mu(q)$ using
the Gordon identity to other operators with other form factors, each of
which are equivalent on shell, but not off shell.  This issue has been
considered by de~Forest~\cite{def}, and by Chinn and Picklesimer~\cite{cpc},
who have given examples of possible choices of extrapolation, and studied
the sensitivity of the response functions and of the Coulomb sum to those
choices.  Our method is similar in principle, but is based specifically
on the Gordon transformation which relates $F_1$ and $F_2$ to the free
Sachs form factors $G_E$ and $G_M$, given in (\ref{bh}) for free nucleons.
We consider two possibilities: that the functional dependence on $Q^2$ of
the functions $F_1(Q^2)$, $F_2(Q^2)$ is unchanged off shell (Model F), or
alternatively, that the functional dependence on $Q^2$ of the functions
$G_E(Q^2)$, $G_M(Q^2)$ is unchanged off shell (Model G).  Other choices
are possible.  For example, in principle, the vector field $V_0$ may also
enter the current operator.  We restrict our attention to Models F and G,
which illustrate the important issues and do not depend on $V_0$.

The first choice, Model F, is the most common; here it is simply assumed
that the Dirac and anomalous form factors $F_1(Q^2)$ and $F_2(Q^2)$ are
unchanged in the nuclear environment, i.e., that they are given by
(\ref{bh}) where both the Sachs form factors and the kinematic variable
$\tau\!\equiv\!Q^2/4M^2$ are evaluated at the actual momentum transfer of
the experiment, i.e., at the momentum
$q_\mu\!\equiv\!k_\mu\!-\!k_\mu^\prime$ found at the electron vertex.
This is in some sense a minimal assumption, in that the
effects of the scalar field enter explicitly the current operator
${\hat J}_\mu(q)$ {\it only} through the nucleon field operators. In
spite of its apparent simplicity, this model has the peculiar feature
that the Dirac and anomalous magnetic moments are treated differently
in the nuclear medium.  This follows since the Dirac moment ($e/2M^*$)
scales with the nucleon mass $M^*$, as can be seen from the solution of
(\ref{bd}) for a nucleon at rest in the presence of a uniform magnetic
field, while the anomalous moment ($\kappa e/2M$) is unchanged in the
medium.  This behavior is possible for a theory with point Dirac nucleons
dressed by charged meson fields, such as QHD-II~\cite{sw}, since the
anomalous moments in such a theory are typically included explicitly by
hand, following a solution of the many-body problem using point nucleons.
However, this behavior is not likely for a theory with internal nucleon
electromagnetic structure, such as QCD, since there one expects that the
Dirac and anomalous magnetic moments may have similar origins, and
therefore may respond in a similar manner to the nuclear environment.

An alternative assumption, which we call Model G, is that the Sachs
electric and magnetic form factors $G_E(Q^2)$ and $G_M(Q^2)$ are unchanged
in the nuclear medium.  To express $\Gamma_\mu(q)$ in the form (\ref{bg}),
the transformation (\ref{bh}) is now performed in the $M^*$-basis: the
off-shell forms of $F_1$ and $F_2$ are now given in terms of the free
Sachs form factors by

	\begin{eqnarray}F_1(Q^2;\tilde\tau^*)
	&=&{G_E(Q^2)+\tilde\tau^* G_M(Q^2)\over1+\tilde\tau^*}\nonumber\\
	& &\label{bi}\\ \kappa F_2(Q^2;\tilde\tau^*)
	&=&{M\over M^*}{G_M(Q^2)-G_E(Q^2)\over1+\tilde\tau^*},
	\nonumber\end{eqnarray}

\noindent where the new variable $\tilde\tau^*\!\equiv\![{\bf q}^2-
(E^*_{{\bf p}+{\bf q}}\!-\!E_{\bf p}^*)^2]/4{M^*}^2$.
In contrast to Model F described above, the off-shell
choice (\ref{bi}) leads to a total nucleon magnetic moment equal to
$(1\!+\!\kappa)\,e/2M^*$.  In particular, the effective Dirac and anomalous
magnetic moments behave similarly in the nuclear medium.  (Note that the
factor $M/M^*$, appearing on the right-hand side of $F_2$ in (\ref{bi}),
has the effect of replacing $\kappa/2M$ by $\kappa/2M^*$ in (\ref{bg}).)
Also, since $\tilde\tau^*$ depends {\it only} on the momenta ${\bf p}$
and ${\bf q}$, the photon energy $\omega$ now enters the form factors
$F_1$ and $F_2$ {\it only} through the Sachs form factors $G_E(Q^2)$
and $G_M(Q^2)$.  This can be seen by expressing $\Gamma_0$ in the
alternate form given in (A5) of Appendix A.  In the next section,
we will see that Model G leads to relativistic Coulomb sum rule which
is the direct analog of that derived in the free PWIA, but with
$M\!\rightarrow\!M^*$.


\section{Coulomb Sum Rules}
\setcounter{equation}{0}
\seceqc

The derivation of a RCSR based on the interacting PWIA of Section 2 is
formally similar to that based on the free PWIA of Ref.~\cite{fk}.
We begin with the Coulomb response function (\ref{bb}), and formally
evaluate the spacelike Coulomb sum (\ref{aa}) by integration over the
photon energy $\omega$.  As described in Ref.~\cite{fk}, to arrive at a
{\em non-energy-weighted} sum rule it must be possible to factor
{\it all} dependence on $\omega$ (which here enters through $Q^2$)
from the current matrix element in (\ref{bb}).  This can be
accomplished by requiring that the ratio $\Gamma_0(q)/G_{Ep}(Q^2)$
be independent of $\omega$, since the plane-wave spinors appearing
in (\ref{bc}) are functions only of the 3-momenta {\bf p} and {\bf q}.
For a system of Dirac protons, for which $\Gamma_0\!=\!\gamma_0$,
this is satisfied trivially.  In Model G, it is satisfied by explicit
construction, using the assumption of proportional Sachs form factors.
For more complicated off-shell models, such as Model F, which do involve
explicit dependence on $\omega$, the derivation of a RCSR may require
further assumptions.  In this article, we will assume that the above
condition is satisfied, and outline the remaining steps which lead
to a RCSR.

We next use closure to perform the sum over final states in the squared
matrix element.  In the nonrelativistic Coulomb sum rule, in which the
integration is over all $\omega$, the use of closure here is exact.
In the relativistic case, the use of closure over the spacelike states
alone requires certain assumptions about the spacelike nuclear excitation
spectrum, as discussed in Ref.~\cite{fk}.  These arguments were based
primarily on the example of a Fermi gas.   We make similar arguments here,
and assume that the spacelike spectrum is saturated by the nucleons-only
response, where now ``nucleons-only'' refers to positive-energy baryons of
mass $M^*$.  After performing the sum over final states $|f\rangle$, the
momentum-space anticommutation relations are used to separate one- and
two-body terms, as in (\ref{ab}).  The relativistic recoil function which
appears in (\ref{ac}) is of the form

        \begin{equation}r_\sigma({\bf p},{\bf q})=
        {1\over2}\sum_{ss^\prime}\ \Bigl\vert
        j_{s^\prime s,\sigma}({\bf p},{\bf q})
        \Bigr\vert^2,\label{rjpq}\end{equation}

\noindent where the matrix element is defined by

	\begin{equation}j_{s^\prime s,\sigma}({\bf p},{\bf q})\equiv
	{{\bar u}_{s^\prime\sigma}({\bf p}\!+\!{\bf q})
	\over\sqrt{2E_{\bf p+q}^*}}
	{\Gamma_0(q)\over G_{Ep}(Q^2)}
	{u_{s\sigma}({\bf p})\over\sqrt{2E_{\bf p}^*}},
	\label{jpq}\end{equation}

\noindent and $u_{s\sigma}({\bf p})$ represents a nucleon with momentum
{\bf p}, spin $s$ and isospin projection $\sigma$.  The above expressions
are identical in form to those obtained in the free PWIA of Ref.~\cite{fk},
but with the replacement $M\!\rightarrow\!M^*$.

Although the one-body term, evaluated using (\ref{rjpq}) and (\ref{jpq})
would account {\it exactly} for the relativistic effects of nucleon recoil
and Fermi motion, it requires knowledge of the nucleon momentum distribution
$n_\sigma({\bf p})$ in the target ground state, which is not normally
known precisely.  We therefore expand the recoil factor
$r_\sigma({\bf p},{\bf q})$ in powers of the nucleon momentum {\bf p},
as in Ref.~\cite{fk}.  Assuming spherical symmetry, the one-body term
takes the form\footnote{We have changed notation slightly from
Ref.~\cite{fk}, in naming the expansion coefficients
$r_{i\sigma}({\bf q})$.}

        \begin{equation}\Sigma^{(1)}({\bf q})=
        \sum_\sigma\ N_\sigma\ r_{i\sigma}({\bf q})
        \ \langle{\bf p}^i\rangle_\sigma~,
        \label{momexp}\end{equation}

\noindent where the momentum moments are defined

        \begin{equation}\langle{\bf p}^i\rangle_\sigma\equiv
	{2\over N_\sigma}\ \sum_{\bf p}\ n_\sigma({\bf p})
	\ {\bf p}^i,\label{momdef}\end{equation}

\noindent and $N_\sigma\!=\!Z,N$ for $\sigma\!=\!p,n$, respectively.
In Ref.~\cite{fk}, we explained that only the first few moments are
required to obtain an accurate result for the one-body term.  In
the next section, we will study how the convergence properties of
this moment expansion depend on the parameter $M^*$, and on the
particular choice of off-shell model.  We now give the precise forms
of the recoil factor $r_{i\sigma}({\bf q})$ in Models F and G.

\subsection{Sum Rules in Model G}

We first consider Model G, in which we assume that the medium
dependence of the nucleon electromagnetic coupling is
described by the off-shell form factors (\ref{bi}).  By construction the
form factors $F_1$ and $F_2$ depend on $Q^2$ only through their linear
dependence on the free Sachs form factors, which are assumed
to be proportional.
This ensures that the ratio $\Gamma_0/G_{Ep}(Q^2)$ is independent of
the photon energy $\omega$.  With this condition satisfied, and since
the baryon spinors are also independent of $\omega$, the matrix element
$j_{s^\prime s,\sigma}({\bf p},{\bf q})$ is a function only of
three-momenta and  the separation of one- and two-body terms in
$\Sigma({\bf q})$ proceeds as described above, leading to (\ref{ac}).

Using the form factors (\ref{bi}) in (\ref{bg}) to evaluate
the matrix element $j_{s^\prime s,\sigma}({\bf p},{\bf q})$ we obtain

	\begin{equation}r_\sigma({\bf p},{\bf q})
	=\epsilon_\sigma^2\ r_E({\bf p},{\bf q})
	+\mu_\sigma^2\ r_M({\bf p},{\bf q}),
    \label{rg}\end{equation}

\noindent where we have defined the electric (E) and magnetic (M)
recoil functions:

        \begin{equation}r_E({\bf p},{\bf q})\equiv
        {1\over1+{\tilde\tau}^*}{(E_{\bf p+q}^*+E_{\bf p}^*)^2
        \over4E_{\bf p}^*E_{\bf p+q}^*},
        \label{repq}\end{equation}

\noindent and

        \begin{eqnarray}r_M({\bf p},{\bf q})
        &\equiv&{1\over1+{\tilde\tau}^*}
        {\tilde\tau^*(E_{\bf p+q}^*+E_{\bf p}^*)^2-(1+{\tilde\tau}^*)
        {\bf q}^2\over4E_{\bf p}^*E_{\bf p+q}^*}\nonumber\\
        &&\label{rmpq}\\
        &=&{1\over1+{\tilde\tau}^*}{1\over4{M^*}^2}
        {|{\bf p}\times{\bf q}|^2\over4E_{\bf p}^*E_{\bf p+q}^*}.
        \nonumber\end{eqnarray}

\noindent The functional forms of $r_E$ and $r_M$ can alternatively
be obtained by writing $\Gamma_\mu$ of (2.7) in terms of $G_E$ and
$G_M$, as explained in Appendix A.
In (\ref{rg}) we have also defined the nucleon electric
charge $\epsilon_\sigma$ (in units of the proton charge) and the
total magnetic moment $\mu_\sigma$ (in nuclear magnetons):

	\begin{eqnarray}
	\epsilon_\sigma &\equiv&{G_{E\sigma}(Q^2)
	\over G_{Ep}(Q^2)}=\biggl\lbrace{1,{\rm\ \ for\ }\sigma=p
	\atop 0,{\rm\ \ for\ }\sigma=n}\nonumber\\
	& &\label{emg}\\
	\mu_\sigma &\equiv&{G_{M\sigma}(Q^2)\over G_{Ep}(Q^2)}
	=\biggl\lbrace{1+\kappa_p,{\rm\ \ for\ }\sigma=p
	\atop \ \ \kappa_n,\ \ \ {\rm\ \ for\ }\sigma=n.}
	\nonumber\end{eqnarray}

\noindent
The functional form of the recoil function (\ref{rg}) is identical to
that obtained in the free PWIA of Ref.~\cite{fk}, but with
$M\!\rightarrow\!M^*$.  This was ensured by choosing to include
the factor $M/M^*$ in (\ref{bi}).  As described above, we expand
$r_\sigma({\bf p},{\bf q})$ in powers of the nucleon momentum
{\bf p},  which with (\ref{ac}), leads to an expression for the
one-body term $\Sigma^{(1)}({\bf q})$, as in (\ref{momexp}).
The expansion coefficients are given in Appendix B,
and are identical to those obtained in the free PWIA of Ref.~\cite{fk},
but with $M\!\rightarrow\!M^*$.  In this model, therefore, we can
include the relativistic effects of a strong scalar field, in a
straightforward extension of the RCSR derived in the free PWIA.

\subsection{Sum Rules in Model F}

We now return to Model F, in which we assume that the nucleon form factors
are given by (\ref{bh}).  Now the photon energy $\omega$ enters not only
through $Q^2$, which appears in the free Sachs form factors, but also
through $\tau\!\equiv\!Q^2/4M^2$, which enters in the transformation
(\ref{bh}).  This additional dependence on $\omega$ can {\it not} be
removed by simply dividing by an overall factor $G_{Ep}(Q^2)$,
as in Model G, without some further approximation.

In order to proceed, we will first calculate the matrix element
appearing in (\ref{rjpq}), using the vertex operator (\ref{bg}) and
the form factors (\ref{bh}).  Since division by $G_{Ep}(Q^2)$ is not
sufficient to render the matrix element a function only of the momenta
{\bf p} and {\bf q}, this function will not yet lead to the one-body
term of the Coulomb sum rule.  However, it does serve as a useful starting
point to illustrate the important issues.  We obtain

        \begin{equation}r_\sigma({\bf p},{\bf q};Q^2)
	=\Biggl[{{G_{E\sigma}^*}(Q^2,\tilde\tau^*)
	\over G_{Ep}(Q^2)}\Biggr]^2\ r_E({\bf p},{\bf q})
	+\Biggl[{{G_{M\sigma}^*}(Q^2,\tilde\tau^*)
	\over G_{Ep}(Q^2)}\Biggr]^2\ r_M({\bf p},{\bf q}),
	\label{rf}\end{equation}

\noindent where we have emphasized explicitly the dependence on $Q^2$.
The form of (\ref{rf}) is similar to (3.5) for Model G, but with
``effective'' Sachs electric and magnetic form factors, defined by:

	\begin{eqnarray}
	G_{E\sigma}^*(Q^2,\tilde\tau^*)&\equiv&F_{1\sigma}(Q^2)
	-\kappa_\sigma{M^*\over M}\tilde\tau^*F_{2\sigma}(Q^2)
	\nonumber\\ &&\label{gsdef}\\
	G_{M\sigma}^*(Q^2,\tilde\tau^*)&\equiv&F_{1\sigma}(Q^2)
	+\kappa_\sigma{M^*\over M}F_{2\sigma}(Q^2),
	\nonumber\end{eqnarray}

\noindent which reduce to the free Sachs form factors if $M^*\!=\!M$.
Inserting $F_1(Q^2)$, $F_2(Q^2)$ of (\ref{bh}) into (\ref{gsdef}),
we have

	\begin{eqnarray}
	{G_{E\sigma}^*(Q^2,\tilde\tau^*)\over G_{Ep}(Q^2)}&=&
	\epsilon_\sigma\ \Biggl[{1+{M^*\over M}\tilde\tau^*\over1
	+\tau}\biggr]+\mu_\sigma\ \Biggl[{\tau-{M^*\over M}\tilde
	\tau^*\over1+\tau}\Biggr]\nonumber\\&&\label{gsofg}\\
	{G_{M\sigma}^*(Q^2,\tilde\tau^*)\over G_{Ep}(Q^2)}&=&
	\epsilon_\sigma\ \Biggl[{1-{M^*\over M}\over1+\tau}\Biggr]
	+\mu_\sigma\ \Biggl[{\tau+{M^*\over M}\over1+\tau}\Biggr],
	\nonumber\end{eqnarray}

\noindent
which illustrates of ``mixing'' of (free) electric and magnetic contributions
in this model.  This mixing enters as a result of the reduced effective
nucleon mass, and is distinct from the usual Lorentz mixing which occurs
for a moving particle.  Effective form factors of the sort (3.11) can
also be seen in Ref.~\cite{ddb}, for example.

To continue the derivation of a {\em non-energy-weighted} sum rule,
we must make some assumption about the remaining $\omega$-dependence,
which arises through the variable $\tau$ in (3.11).  A simple method is
based on the excitation energy of a uniform Fermi gas in QHD-I,
which is expressible simply in terms of the three-momenta {\bf p}
and {\bf q}: $\omega\!=\!E_{\bf p+q}^*\!-\!E_{\bf p}^*$.
We therefore make the replacement in (3.11):

	\begin{equation}\tau(Q^2)\rightarrow\tau(Q^2)\Bigr
	\vert_{\omega=E_{\bf p+q}^*-E_{\bf p}^*}
	\ ={\tilde\tau}^*\biggl({M^*\over M}\biggr)^2,
	\label{tau}\end{equation}

\noindent to obtain energy-independent effective form factors.
The factor $(M^*/M)^2$ arises directly from the definition of $\tau$
following (2.8).  With the replacement (\ref{tau}) in (3.11), the
recoil function (3.9) is now independent of the photon energy $\omega$,
and the derivation of a non-energy-weighted Coulomb sum rule for Model F
proceeds as for Model G.

At this point, it is possible to expand (3.9) about ${\bf p}\!=\!0$
and obtain a moment expansion for $\Sigma^{(1)}({\bf q})$, as in
(\ref{momexp}).  However, the resulting coefficients $r_{i\sigma}({\bf q})$
are much more complicated than those which arise from (3.5) in Model G,
due to the dependence on {\bf p} of the effective form factors
(3.11).  In addition, we will see that the essential physics
of Fermi motion enters (\ref{rf}) only through the electric and magnetic
recoil functions $r_E({\bf p},{\bf q})$ and $r_M({\bf p},{\bf q})$, and
can be ignored in the effective form factors themselves.  It is therefore
useful to consider an approximation scheme which will allow a more
efficient evaluation of the one-body term: setting ${\bf p}\!=\!0$
in (3.11) after applying (3.12), we can write

	\begin{equation}r_\sigma({\bf p},{\bf q})
	={\epsilon_\sigma^*}^2({\bf q})\ r_E({\bf p},{\bf q})
	+{\mu_\sigma^*}^2({\bf q})\ r_M({\bf p},{\bf q}),
	\label{rfnew}\end{equation}

\noindent where we have defined the ``effective'' nucleon charges and
magnetic moments:

    \begin{eqnarray}
	\epsilon_\sigma^*({\bf q})&\equiv&
	{G_{E\sigma}^*(Q^2,\tilde\tau^*)\over G_{Ep}(Q^2)}
	\Biggr\vert_{{\bf p}=0}\nonumber\\
	&&\nonumber\\
	&=&\epsilon_\sigma\ \Biggl[{2M^2+M(E_{\bf q}^*-M^*)
	\over 2M^2+M^*(E_{\bf q}^*-M^*)}\Biggr]
	+\mu_\sigma\ \Biggl[{(M^*-M)(E_{\bf q}^*-M^*)
	\over 2M^2+M^*(E_{\bf q}^*-M^*)}\Biggr]
	\nonumber\\ & &\label{emf}\\
	\mu_\sigma^*({\bf q})&\equiv&
	{G_{M\sigma}^*(Q^2,\tilde\tau^*)\over G_{Ep}(Q^2)}
	\Biggr\vert_{{\bf p}=0}\nonumber\\
	&&\nonumber\\
	&=&\epsilon_\sigma\ \Biggl[{2M(M-M^*)
	\over 2M^2+M^*(E_{\bf q}^*-M^*)}\Biggr]
	+\mu_\sigma\ \Biggl[{2MM^*+M^*(E_{\bf q}^*-M^*)
	\over 2M^2+M^*(E_{\bf q}^*-M^*)}\Biggr],
	\nonumber\end{eqnarray}

\noindent following (\ref{emg}).  We will refer to expression (\ref{rfnew})
as the ``factored moment expansion.''  The effective form factors now
enter (\ref{rfnew}) as functions only of the momentum transfer {\bf q},
but do not complicate its expansion in {\bf p}.  That this simplification
is a valid approximation will be shown numerically in the next section.


\section{Sum Rules for Relativistic Nuclear Matter}
\setcounter{equation}{0}
\seceqd

We now apply the sum rules derived in the previous section to a model
nuclear system: uniform nuclear matter in the relativistic mean field
approximation, as given by QHD-I.~\cite{sw}  For this system the effects
of nuclear structure on the Coulomb sum $\Sigma({\bf q})$ enter
through the effective nucleon mass $M^*$ and through the nuclear
electromagnetic current, as we have discussed in Section 3.
We are interested in the dependence of the Coulomb sum on $M^*$
and on the choice of off-shell model (F or G) for the current.
We then demonstrate how to apply the sum rule methods of Section 3
to the model system, as if it were measured Coulomb response data.
The analysis will necessarily be model dependent, through the choice
of $M^*$ and the off-shell model (F or G), as well as through the
moments $\langle{\bf p}^i\rangle_\sigma$ of the nucleon momenta in
the target.  We shall examine the accuracy of the moment expansion
for the ``best-case'' analysis, that is, for which the sum rule
parameters match those of the assumed nuclear model.  This will
illustrate the convergence properties of the moment expansion
in each model, and the validity of the factored moment expansion
in Model F.

\subsection{Test in Fermi Gas Model}

We begin by computing the Coulomb sum $\Sigma({\bf q})$ for a uniform
Fermi gas of nucleons moving in uniform vector and scalar potentials.
This has been studied previously~\cite{mat,ddb}.  The result can be
written in the form

        \begin{equation}\Sigma({\bf q})=\int{d^3p\over(2\pi)^3}
        \ \sum_\sigma n_\sigma({\bf p})
        \bigl[1\!-\!n_\sigma({\bf p}\!+\!{\bf q})\bigr]
        \ \sum_{ss^\prime}\ \Bigl\vert
        j_{s^\prime s,\sigma}({\bf p},{\bf q})
        \Bigr\vert^2,\label{rcsfg}\end{equation}

\noindent where the matrix element $j_{s^\prime s,\sigma}({\bf p},{\bf q})$
is defined in (\ref{jpq}).  We will evaluate (\ref{rcsfg}) by numerical
integration over $p$ and $\theta$, the angle between {\bf p} and
{\bf q}.  The result is independent of $V_0$, for reasons given
following (\ref{be}). The factor $n_\sigma({\bf p})\!\equiv\!\theta
(p_{F\sigma}\!-\!|{\bf p}|)$ restricts the sum over initial states to
include only those states which are occupied in the nuclear ground state,
and the factor $\bigl[1\!-\!n_\sigma({\bf p}\!+\!{\bf q})\bigr]$ ensures
that final states for which $|{\bf p}\!+\!{\bf q}|\!\le\!p_{F\sigma}$
are excluded.  This Pauli exclusion is the only source of two-particle
correlations in this simple model.

We shall consider two particular values of $M^*$ which arise in QHD-I,
as well as $M^*\!=\!M$. In the mean-field theory (MFT), static polarization
of the baryon vacuum in the nuclear ground state (due to the scalar field
$\phi$) is ignored in the coupled field equations.  In the relativistic
Hartree approximation (RHA), static vacuum polarization is included at
the one-loop level.\footnote{``MFT'' and ``RHA'' are used here in the
conventional sense defined in Ref.~\cite{sw}.} The particular values of
$M^*$ are obtained from the self-consistent solutions for uniform nuclear
matter at saturation.  Taking $Z\!=\!N$ and using the input parameters
$p_F\!=\!1.42{\rm\ fm}^{-1}$ and ${\rm BE/N}\!=\!-15.75{\rm\ MeV/nucleon}$,
leads to the values $M^*/M\!=\!0.556$ in MFT, and $M^*/M\!=\!0.718$ in RHA.
Although it may be argued that the model assumptions of MFT are technically
more consistent with the nucleons-only approximation, it may also be true
that the resulting effective mass $M^*$ is too small due to the
simplicity of the model, as compared to optical potential
phenomenology (see, e.g., Ref.~\cite{wal}).  In the calculations to
follow, we will show results using both the MFT and RHA values of $M^*$
to illustrate the kinds of effects which can be expected in this range.

For applications to data, it is convenient to cast the RCSR in a slightly
different form, so that the saturation of the sum rule is more apparent.
As in Ref.~\cite{fk}, we define the ``modified'' Coulomb sum
\footnote{Note the factor $1/Z$, which is not present in $S_A$ as
defined in Ref.~\cite{fk}.}

        \begin{equation}S_{A}({\bf q})\equiv{1\over Z}
        {\Sigma({\bf q})\over r_{A}({\bf q})},
        \qquad A=I,II,III\ldots\label{sdef}
        \end{equation}

\noindent where the recoil ``correction'' factors are defined

        \begin{equation}r_{I}({\bf q})\equiv\biggl[r_{0p}+{N\over Z}
        r_{0n}\biggr]\label{ridef}\end{equation}

        \begin{equation}r_{II}({\bf q})\equiv r_{I}({\bf q}) +
        \biggl[r_{2p}\langle{\bf p}^2\rangle_p
        +{N\over Z}r_{2n}\langle{\bf p}^2\rangle_n\biggr],
        \label{riidef}\end{equation}

\noindent and so on for higher orders.
\footnote{In this notational scheme, $r_A({\bf q})$ includes terms through
${\cal O}({\bf p}^{2A-2})$.}  The expansion coefficients $r_{i\sigma}
({\bf q})$ are given in Appendix B.  The moments $\langle{\bf p}^i
\rangle_\sigma$ are defined in (3.4).  For applications of the RCSR to
$\Sigma({\bf q})$ of (\ref{rcsfg}), we will assume the sharp distribution
$n_\sigma({\bf p})\!=\!\theta(p_{F\sigma}-|{\bf p}|)$ for the evaluation
of the momentum moments.  Integrating this distribution over the Fermi
sphere of radius $p_{F\sigma}$ leads to the $i^{\rm th}$ moment, given
by
        \begin{equation}\langle{\bf p}^i\rangle_\sigma
        ={3\over3+i}\ p_{F\sigma}^i~.\label{momfg}\end{equation}

\noindent This is consistent with the PWIA and the evaluation of the
RCSR on (\ref{rcsfg}).  For applications to actual data, however, it
would be preferable to use the most accurate values available for
these moments, e.g., using experimentally determined $n_\sigma({\bf p})$
from $(e,e^\prime p)$.

Following (\ref{ab}), the above definitions lead to an {\em approximate}
sum rule of the form

        \begin{equation}S_A({\bf q})\simeq 1+
	{1\over Z}{1\over r_A({\bf q})}\
	\biggl[C({\bf q})+\Sigma_{\rm un}^{(2)}({\bf q})
	\biggr],\label{sapp}\end{equation}

\noindent as discussed in Section 5 of Ref.~\cite{fk}.  The Fermi gas model
includes
only Pauli correlations, which vanish identically for $q\!\ge\!2p_F$.
Consequently, an application of the RCSR to (\ref{rcsfg}) will be considered
``accurate'' if the modified Coulomb sum $S_A({\bf q})$ tends to unity for
momenta beyond the range of these correlations, i.e., if $S_A({\bf q})\!
\rightarrow\!1$ for $q\!\ge\!2p_F$.  This is frequently referred to in the
literature as ``saturation'' of the sum rule.

\subsection{Numerical Results}

In the following examples, we take $Z\!=\!N$.
To demonstrate the importance of the effects of a reduced effective
mass $M^*$, we first consider a system of Dirac protons, for which
$\epsilon_p\!=\!\mu_p\!=\!1$ and $\epsilon_n\!=\!\mu_n\!=\!0$.
In Ref.~\cite{fk}, we showed for $M^*/M\!=\!1$ that the lowest-order
sum rule, $S_I({\bf q})$, is accurate to within $\sim\!1\%$ for this case
(see Figure 2 in Ref.~\cite{fk}).  In Figure 1, the dotted and dot-dashed
curves show the Coulomb sum $\Sigma({\bf q})$ calculated with $M^*/M\!=\!1$
and $M^*/M\!=\!0.556$, respectively.  In this example, we see a 10--20\%
reduction in the Coulomb sum over this momentum range, due to the reduced
effective mass $M^*$.  This effect has been noted previously for
lower momenta~\cite{mat,ddb}.  The dashed and solid curves show the lowest
order sum rule, $S_I({\bf q})$, applied to $\Sigma({\bf q},M^*)$ in two
different ways:  In the dashed curve, the recoil factor $r_I({\bf q})$
has been evaluated at the free nucleon mass $M$, while the solid curve
shows a consistent application of the sum rule, in which $r_I({\bf q})$
has been evaluated at the reduced mass $M^*$.  The dashed curve
demonstrates that if the effective nucleon mass is substantially reduced
in the medium, then one must account for this in the application of the
RCSR.  The solid curve demonstrates that even when the effective mass is
greatly reduced, e.g., $M^*/M\!=\!0.556$, the lowest-order sum rule
$S_I({\bf q})$ applied to a system of Dirac nucleons is accurate to
within $\sim\!1\%$, {\em if} the appropriate value of $M^*$ is used.

In Figure 2, we show the RCSR applied to a uniform system of nucleons
with anomalous magnetic moments, for $M^*/M\!=\!1$.  (Here Models F
and G are identical.)  A similar figure was shown in Ref.~\cite{fk},
but it is useful to review the main results here for comparison.
The unmodified Coulomb sum $\Sigma({\bf q})$ is shown by the dotted curve,
and indicates an enhancement relative to the free Dirac result. This is a
consequence of the nucleon anomalous moments, as explained in Ref.~\cite{fk}.
The lowest-order sum rule, $S_I({\bf q})$, overshoots the
``correct'' result, i.e., saturation, increasingly with higher momentum,
and indicates the importance of Fermi motion effects when anomalous moments
are included.  In Ref.~\cite{fk} we explained that this is because the
magnetic contribution is enhanced relative to the electric by a factor
$\mu_p^2\!+\!\mu_n^2\!\simeq\!11.4$.  This prevents the cancellation of
electric and magnetic Fermi motion effects, which occurs nearly exactly
for Dirac nucleons.  The striking feature is that the ${\cal O}({\bf p}^2)$
result, $S_{II}({\bf q})$ is accurate to within $\sim 1\%$ over this
momentum range.  We also show the ${\cal O}({\bf p}^4)$ result,
$S_{III}({\bf q})$, which is accurate to better than $0.1\%$ over the
same range. This we interpret as numerical evidence of convergence of the
moment expansion in this model.  The results shown here will be modified
for $M^*/M\!<\!1$, and will also depend on the choice of form factor model
(F or G).  We are interested in the accuracy of sum rules designed to
account for these modifications.

We now consider $M^*/M\!<\!1$.  In Figures 3a and 3b we show results
for the RCSR in Model G for RHA and MFT values of $M^*$, respectively.
As for Figure 2, the Coulomb $\Sigma({\bf q})$ is enhanced relative to
that for Dirac nucleons, due to Fermi motion effects, when anomalous
moments are included.  This model also has the rather peculiar feature
that for $M^*/M\!\simeq\!1/2$, we have $\Sigma({\bf q})\!\simeq\!1$ for
$q\!>\!2p_F$.  This value is coincidental (see Sec.~4.3 for further
discussion).  For still lower values of $M^*$ the Coulomb sum can be
significantly greater than unity.  This is contrary to the common notion
that relativistic effects necessarily suppress the Coulomb sum relative
to unity.  Note that the moment expansion converges more slowly for
smaller $M^*$, as can be seen by comparing $S_{II}({\bf q})$ in Figures
3a and 3b.  Still, the RCSR to ${\cal O}({\bf p}^2)$ is reasonably accurate,
even for $M^*/M\!\sim\!1/2$.  The very high accuracy of $S_{III}({\bf q})$
is interpreted as numerical evidence of convergence in this model.
Indeed, for $q\!>\!2p_F$, $S_{III}\!=\!1.00$ for the RHA mass and
$S_{III}\!=\!1.01$ for the MFT mass, at $q\!=\!4{\rm\ GeV}$.

In Figures 4a and 4b we show analogous results for a nuclear system
under the off-shell assumption of Model F.   The Coulomb sum
$\Sigma({\bf q})$ behaves qualitatively differently in this model, with
a minimum developing around $q\!\sim\!2{\rm\ GeV}$ as $M^*$
decreases to the MFT value.  The differences originate from the
additional $q$-dependence in the effective form factors
$G_{E\sigma}^*/G_{Ep}$ and $G_{M\sigma}^*/G_{Ep}$, which are
constants in Model G.  As in Model G, $S_I({\bf q})$ is not adequate,
and illustrates the importance of Fermi motion effects when anomalous
moments are included.  Also as for Model G, $S_{II}({\bf q})$, shown by
the dashed curve, is reasonably accurate over this range of momenta and
$M^*$.  The improvement in $S_{III}({\bf q})$ is interpreted as numerical
evidence of convergence of the moment expansion.  The most important
feature of the sum rule results just presented for Model F is the success
of the factored moment expansion, which allows the use of the {\it same}
expansion coefficients as in Model G, i.e., those given in Appendix B.
This approximation neglects only the Fermi motion effects which enter
through the ratios $[G^*/G_{Ep}]^2$ in (3.9).
To illustrate its accuracy, we have evaluated $S_{II}({\bf q})$
in a ``consistent'' moment expansion which correctly includes to
${\cal O}({\bf p}^2)$ additional Fermi motion effects entering
through these ratios.  This result is shown
by a dot-dashed curve, which is nearly identical to the ``factored''
result, shown by the dashed curve.  These are small corrections on the
scale of the other kinematical effects, lending support to the factored
moment expansion in Model F.

In summary, we have shown that the relativistic effects of nucleon
recoil, Fermi motion and a reduced effective mass $M^*$ can be accounted
for accurately in both Models F and G.  Keeping only terms through
${\cal O}({\bf p}^2)$ gives $\sim\!5\%$ accuracy, which is small on
the scale of the dominant kinematical effects.  Keeping terms
through ${\cal O}({\bf p}^4)$ gives $\sim\!1\%$ accuracy.
We emphasize that in both Models F and G, the RCSR was evaluated
using the {\em same} moment expansion coefficients (those in Appendix
B), but in Model F the effective charges and magnetic moments depend on
$q$ and $M^*$.  This simplification requires using a factored moment
expansion, in which Fermi motion effects entering though the effective
form factors $G_{E\sigma}^*$, $G_{M\sigma}^*$ are ignored.  In the cases
studied this is an excellent approximation, as explained further below.

\subsection{Discussion of Results}

One can understand the results just presented by looking individually
at the electric and magnetic recoil functions $r_E$ and $r_M$, defined
in (\ref{repq}) and (\ref{rmpq}), as functions of $M^*$.  We shall
look at the variations of $\epsilon_\sigma^*({\bf q})$ and $\mu_\sigma^*
({\bf q})$ of (3.14) in Model F, compared to $\epsilon_\sigma$ and
$\mu_\sigma$ in Model G.  We are especially interested in the factored
moment expansion in Model F.  We  shall examine the reasons for its
success, and under what conditions such an approach may be expected
to work for other off-shell assumptions.  We argue that a factored moment
expansion, like that adopted here, should be accurate in a large class
of off-shell form factor models.

In Figures 5a and 5b, we show the electric and magnetic recoil functions:
$r_E({\bf p},{\bf q})$ of (\ref{repq}) and $r_M({\bf p},{\bf q})$ of
(\ref{rmpq}), for two values of $M^*$.  In both figures, the solid curve
represents the electric term $r_E({\bf 0},{\bf q})$, which dominates
the Coulomb response for a system of Dirac nucleons.  The magnetic
term $r_M({\bf 0},{\bf q})\!=\!0$, as can be seen from the second equality
in (\ref{rmpq}).  Thus the Coulomb response for a nucleon at rest is
purely electric, as noted in Ref.~\cite{fk}.  The electric function
$r_E({\bf 0},{\bf q})$ decreases from unity at $q\!=\!0$ to its limiting
value $1/2$ as $q\!\rightarrow\!\infty$.  These limits are independent of
$M^*$, however, for smaller $M^*$, $r_E({\bf 0},{\bf q})$ decreases more
rapidly with $q$, since it is a monotonic function of $q/M^*$.  This
increased suppression is easily accommodated by using an appropriate
value for $M^*$, as discussed for the Dirac case in Figure 1.

The corrections for Fermi motion to $r_E$ and $r_M$ are also shown
in Figures 5a and 5b, and are denoted by $\delta r_E({\bf q})$ and
$\delta r_M({\bf q})$, respectively.  These corrections were calculated
by averaging $r_E({\bf p},{\bf q})$ and $r_M({\bf p},{\bf q})$ over
the Fermi sphere, taking $p_F\!=\!1.42{\rm\ fm}^{-1}$.
For a system of Dirac nucleons, $\delta r_E$ and $\delta r_M$ enter
the response with equal weighting.  We noted in Ref.~\cite{fk} for
$M^*/M\!=\!1$ that this leads to a nearly exact cancellation over the
entire momentum range, and ensures that the lowest-order sum rule,
RCSR-I, is accurate to within $\sim\!1\%$.  Note that this
cancellation persists to $\sim\!1\%$ accuracy for $M^*/M\!<\!1$,
although {\em both} $\delta r_E$ and $\delta r_M$ nearly triple for
$M^*/M\!\sim\!1/2$, as in the MFT example.

For nucleons with anomalous magnetic moments, this near cancellation of
$\delta r_E({\bf q})$ and $\delta r_M({\bf q})$  no longer occurs,
since these function are multiplied by different factors:
$\epsilon_\sigma^2$, $\mu_\sigma^2$ of (3.8) for Model G, or
${\epsilon_\sigma^*}^2({\bf q})$, ${\mu_\sigma^*}^2({\bf q})$ of
(3.14) for Model F.
In Model G, $\delta r_M({\bf q})$ is multiplied by the constant
factor $\mu_p^2\!+\!\mu_n^2\!\simeq\!11.4$, while
$\epsilon_p^2\!+\!\epsilon_n^2\!=\!1$.  For $M^*/M\!=\!1$, this gives
a 10--20\% increase in the Coulomb sum $\Sigma({\bf q})$, as
was discussed in Ref.~\cite{fk}.  As $M^*/M$ decreases, the enhanced
effect of $\delta r_M({\bf q})$ increases further:  for $M^*/M
\!=\!0.556$, the Fermi motion correction increases with $q$ about as
quickly as $r_E({\bf 0},{\bf q})$ decreases, as shown in Fig.~5b.
This leads to the behavior of $\Sigma({\bf q})$ for this value of $M^*$,
shown in Fig.~3b, and remarked on earlier, i.e., $\Sigma({\bf q})
\!\simeq\!1$.  As $M^*$ is decreased to the MFT value, this increase
in $\delta r_M$ effectively triples the importance of higher
moments.  This accounts for the decreased accuracy of the RCSR
at ${\cal O}({\bf p}^2)$, compared to that applied to free nucleons.

Model F has qualitatively similar behavior, as we saw in Figs.~4a and 4b.
Here, however, $\Sigma({\bf q})$ is modified by the effective charges and
moments (3.14), which are shown in Fig.~6 (solid curves).  For this case
$({\mu_p^*}^2\!+\!{\mu_n^*}^2)$ is somewhat smaller than
$(\mu_p^2\!+\!\mu_n^2)$ and increases with $q$ over the range shown.
This explains the smaller values of $\Sigma({\bf q})$ in Figs.~4a and
4b than in Figs.~3a and 3b, and the upturn with increasing $q$ in Fig.~4b.

We have seen that the convergence properties of the RCSR in Model F are
similar to those in Model G, in spite of the use of a factored moment
expansion that ignores certain Fermi motion effects, which would enter
through the effective form factors (3.11).  In order to understand
this result, we first examine the effective form factors themselves.
In Figure 6, we show these in the form in which they enter the recoil
function $r_\sigma({\bf p},{\bf q};Q^2)$ in Model F, i.e.,
$[G_{E\sigma}^*/G_{Ep}]^2$ and $[G_{M\sigma}^*/G_{Ep}]^2$.
We have used (3.12) and $M^*/M\!=\!0.556$ in their evaluation.
The dashed curves show the form factors at ${\bf p}\!=\!0$,
and the solid curves were obtained by averaging over the Fermi sphere with
$p_F\!=\!1.42{\rm\ fm}^{-1}$.  (Note that the electric form factors tend to
their free values in the limit $q\!\rightarrow\!0$, while the magnetic
form factors tend to their free values in the limit
$q\!\rightarrow\!\infty$.  This can be seen from the definitions given
in (\ref{gsofg}).)  We see that the Fermi motion corrections to the
ratios $[G^*/G_{Ep}]^2$ are small compared to the ratios themselves.
Thus terms ${\cal O}({\bf p}^2)$ and higher are small, especially in
the magnetic form factors.  (Terms ${\cal O}({\bf p}\cdot{\bf q})$
vanish upon angle-integration with spherical symmetry, and therefore
do not contribute to the results in Figure 6.)

To understand the success of the factored moment expansion, consider
the second-order sum rule, $S_{II}({\bf q})$.  It omits two
types of Fermi corrections: 1) ${\cal O}({\bf p}^2)$ terms which come
from a moment expansion of the effective form factors and multiply
$r_E({\bf 0},{\bf q})$ and $r_M({\bf 0},{\bf q})$, and 2) products of
linear terms involving $({\bf p}\cdot{\bf q})$, one term from the effective
form factors, and one from the recoil factors $r_E$ and $r_M$.
The first of these make at most $\sim\!5\%$ correction to the electric
contribution, and less than $\sim\!1\%$ correction to the magnetic,
as can be seen from Figure 6.  These terms can therefore be neglected
to a good approximation.  Linear terms do not enter the magnetic
contribution, as can be seen from the $({\bf p}\times{\bf q})^2$
factor in (\ref{rmpq}).  Linear terms have been omitted from the
electric contribution, and could in principle have been appreciable.
In fact, these terms taken together result in less than a
$1\%$ error in the RCSR, as can be seen by comparing the dashed
and dot-dashed curves for $S_{II}({\bf q})$ in Figures 4a and 4b.
Thus the success of the factored moment expansion in Model F can be
attributed to the smallness of Fermi corrections in the effective form
factors themselves, and in part to the complete absence of linear terms
in the magnetic contribution.

We believe that the success of the factored moment expansion in Model F
may also occur for a much wider variety of off-shell models than we have
considered here.  The basis for this claim is that Fermi corrections tend
to be small in general, representing effects of order $(p_F/M^*)^2$.
The main exception is the magnetic correction, which is substantial
because of two properties: $r_M(0,{\bf q})\!=\!0$, and $({G_{Mp}^*}
^2\!+\!{G_{Mn}^*}^2)/{G_{Ep}}^2\sim10$.  The magnetic contribution
$\delta r_M$ is small, but is then enhanced by roughly an order of
magnitude by the anomalous magnetic moments.  No such enhancement
occurs for $\delta r_E$.  Thus in Model F, the further corrections
due to {\bf p}-dependence in $[G_{E\sigma}^*/G_{Ep}]^2$ and
$[G_{M\sigma}^*/G_{Ep}]^2$ are negligible.  In other off-shell
models, for which $r_\sigma({\bf p},{\bf q})$ is expressible in
the form (3.9), we expect Fermi motion effects in the corresponding
effective form factors also to be small.  We therefore propose
that a factored moment expansion, like that employed here in Model
F, may also be accurate for the evaluation of Fermi motion effects
a wide variety of off-shell models.  This allows one to evaluate the
RCSR using the expansion coefficients given in Appendix B.  (We include
$\delta r_E({\bf q})$, although a small contribution in nucleon models
with anomalous magnetic moments, to ensure the cancellation of Fermi
corrections for the Dirac system, and to allow the ``consistent''
calculation of $S_{II}({\bf q})$ in Figs.~5a and 5b.)


\section{Summary and Conclusions}
\setcounter{equation}{0}
\seceqe

The main results of this paper can be summarized as follows:  The effects
of nuclear binding in the mean-field approximation enter the Coulomb
response function $W_C(\omega,{\bf q})$ and the Coulomb sum
$\Sigma({\bf q})$ in a way that can be characterized by a reduced
effective mass $M^*$.  These effects depend on the behavior of the
electromagnetic current for the off-shell kinematics in the nuclear
medium.  The sensitivity of the Coulomb sum $\Sigma({\bf q})$ to the
choice of off-shell model has been discussed previously~\cite{def,cpc}.
We consider two illustrative models: F and G.  We demonstrate in these
models that the RCSR of Ref.~\cite{fk} can be extended to account for
relativistic binding effects, in addition to the purely kinematic effects
of recoil and Fermi motion treated in Ref.~\cite{fk}.  The resulting
RCSR is no longer model-independent, since one must make some assumptions
about $M^*$ and the off-shell behavior of the form factors.  However,
to the extent that the dominant effects of the nuclear medium can
be characterized by the simple parameter $M^*$, the Coulomb sum rule
analysis can be applied to data, with the goal of looking for nuclear
structure effects beyond the mean field, e.g., two-body correlations.
We emphasize that the form of the sum rule is sufficiently general to
accommodate a broad class of off-shell form factor models, i.e., those
for which the recoil function $r_\sigma({\bf p},{\bf q})$ can be
expressed in the form (3.9).

The real change in transforming to the $M^*$ basis enters through the
nucleons-only approximation, under which the sum rule is derived.
Strong potentials in relativistic models mix free ${\bar N}$ components
into the initial and final interacting nuclear states, and modify the
spacelike response function.  Chinn, Picklesimer and Van
Orden~\cite{cpa,cpb} have isolated the effects of this mixing of
${\bar N}$ components, and show results qualitatively similar to those
seen in Fig.~4.  (Their form factors are closer to our Model F than G.)
The transformation to an $M^*$ basis automatically incorporates these
potential effects in a convenient representation, although the use of
an effective mass independent of position or momentum may only
approximate the physical situation.

How might one apply this RCSR to experimental data?  First, the Coulomb
sum $\Sigma({\bf q})$ must be calculated from the measured Coulomb
response function, as in (\ref{aa}).  Then, one must make some specific
assumptions about the off-shell behavior of the current operator, as
discussed in Section 3.  This should be cast in the $G^*$ form as in
(3.10) and (3.11) for Model F (or $G^*\!\equiv\!G(Q^2)$ for Model G),
with the substitution (3.12) to remove (approximately) any residual
$\omega$-dependence from $G_{E\sigma}^*/G_{Ep}$ and
$G_{M\sigma}^*/G_{Ep}$.  The choice of model is not
restricted to those presented in Sections 3.1 and 3.2.  With the
factored moment expansion, the relativistic recoil function (3.1)
takes the form (3.13) and the ratios $G^*/G_{Ep}$ are evaluated at
${\bf p}\!=\!0$, as in (3.14).  The recoil factor $r_A({\bf q})$ is
then evaluated in a moment expansion, e.g., to ${\cal O}({\bf p}^2)$
as in (4.4), with $\langle{\bf p}^2\rangle_\sigma$ fixed by other
experimental information, or by a model as in (4.5), or as a free
parameter.  The expansion coefficients $r_{i\sigma}({\bf q})$ are
given in Appendix B, and are functions of the parameter $M^*$.

The modified Coulomb sum $S_A({\bf q})$ is obtained by forming the ratio
(4.2) of the experimentally determined numerator $\Sigma({\bf q})$, and
the (model dependent) recoil denominator $Zr_A({\bf q})$.  The expected
behavior of $\Sigma({\bf q})$ with increasing $q$ is that it will approach
the one-body term $\Sigma^{(1)}({\bf q})$ of (1.3), assuming that both
$C({\bf q})$ and $\Sigma_{\rm un}^{(2)}({\bf q})\!\rightarrow\!0$ in
(1.2), as $q\!\rightarrow\!\infty$.  The modified sum $S_A({\bf q})$
will then approach unity {\em if} the assumed form of the current is
correct, {\em and if} the effective mass $M^*$ is appropriately chosen.
Should that be the case for a given set of data, it would be reasonable
to assume that the recoil functions have been correctly chosen.  The sum
rule (1.3) can then be used to investigate the two-nucleon correlation
function in the ratio form

        \begin{equation}S_A({\bf q})\simeq1+{\tilde C}_A({\bf q})
        +{\Sigma^{(2)}_{\rm un}({\bf q})\over Zr_A({\bf q})}
        ,\label{ea}\end{equation}

\noindent by looking for deviations from unity at moderate momentum
transfers.  The two-body correlation function,

        \begin{equation}{\tilde C}_A({\bf q})\equiv
	{1\over Z}{C({\bf q})\over r_A({\bf q})},
        \label{eb}\end{equation}

\noindent is related to the standard nonrelativistic correlation function.
(See Eqs.~(5.8) and (5.16) in Ref.~\cite{fk}.)  An interesting application
of the RCSR is the constraint of the viable off-shell form factor models
by analyzing the experimentally determined Coulomb sum $\Sigma({\bf q})$
in different off-shell models and comparing to $S_A({\bf q})\!\simeq\!1$
beyond the expected range of correlations.

A further remark about the analysis of data with the RCSR seems appropriate.
It has become customary for the experimental Coulomb response data to be
integrated in a modified form of (1.1), as suggested by DeForest~\cite{defg},
in which the proton electric form factor $G_{Ep}(Q^2)$ is replaced by
${\bar G}_{Ep}(Q^2)\!\equiv\!G_{Ep}(Q^2)\sqrt{(1\!+\!\tau)/(1\!+\!2\tau)}$,
with $\tau\!\equiv\!Q^2/4M^2$ (see, e.g., Refs.~\cite{chen,mez}).
We explained in Ref.~\cite{fk} why this procedure will not lead to
a {\em non-energy-weighted} sum rule: to obtain a sum rule of the form
(1.2) or (1.5), the extra $\omega$-dependence should not be introduced
into the definition of the Coulomb sum.  The kinematic effects
included in the recoil factors $r_A({\bf q})$ depend on the 3-momentum
transfer {\bf q}, rather than on the invariant $Q^2$.  It is this property
which preserves the sum rule in passing from (1.2) to (5.1).

The use of the $M^*$-basis for both initial and final states, with the
same value of $M^*$, implicitly assumes that the nucleus is large enough
that the nucleon kinematics in the final state are sensitive to the scalar
field.  For smaller nuclei, it may be necessary to account for the effects
of finite size.  This could possibly be accomplished through a rederivation
of the RCSR in a Hartree representation based on bound, localized nuclear
states.  Another issue is the momentum-dependence of the mean-field
potentials, as discussed in Refs.~\cite{dda,kim}, for example.
Since the RCSR has been derived in momentum space, such a modification
is relatively straightforward.  As a first approximation, which is
consistent with our conclusion that Fermi motion effects tend to be small,
it seems reasonable to use $M^*(0)$ for initial states and $M^*({\bf q})$
for final states.  This prescription preserves the basic structure of the
RCSR, in that no further dependence on ${\bf p}$ is introduced.  This would
require the initial- and final-state masses to be treated distinctly,
however, and would complicate the form of the coefficients in Appendix B.

We are currently investigating the effects of virtual $N{\bar N}$ pairs
(of mass $M^*$) on the RCSR, as they enter intermediate excited states in
the random phase approximation.  These were considered previously by
Horowitz~\cite{hor}, and appear to be significant.  We are also interested
in how energy dependent terms, which can enter in off-shell models where
a substitution of the form (3.12) is not appropriate, affect the RCSR.


\vskip 0.3 true in
\noindent{\Large\bf Acknowledgements}
\vskip 0.2 true in

This research was supported in part by the U.S.~Department of Energy under
Grant No.~DE-FG02-88ER40425 with the University of Rochester.
The authors would also like to thank the High Energy Physics Group
for use of the VAX computer.


\vskip 0.3 true in
\noindent{\Large\bf Appendix A:\ \ \ $\Gamma_\mu$ in Terms of Sachs Form
Factors}
\vskip 0.2 true in
\setcounter{equation}{0}
\seceqaa

An alternative form of the current operator (2.7) may be obtained by
making a Gordon transformation~\cite{bar,han} on matrix elements between
free plane-wave spinors:

	\begin{equation}\Gamma_\mu={1\over1+\tau}\Biggl[
	G_E(Q^2)\ {P_\mu\over M} + G_M(Q^2)\ {r_\mu\over4M^2}
	\Biggr],\label{aaa}\end{equation}

\noindent where $P_\mu\!\equiv\!{1\over2}(p\!+\!p^\prime)_\mu$, and

	\begin{equation}r_\mu\equiv{1\over2}\Biggl[\gamma_\mu
	(P\cdot\gamma)(q\cdot\gamma)-(q\cdot\gamma)(P\cdot\gamma)
	\gamma_\mu\Biggr],\label{aab}\end{equation}

\noindent and the $F$'s and $G$'s are related by (2.8).  Another form of
$r_\mu$ may be obtained~\cite{sca}:

	\begin{equation}r_\mu=2\gamma_5 \epsilon_{\mu\nu\rho\sigma}
	P^\nu q^\rho \gamma^\sigma.\label{aac}\end{equation}

\noindent For $\mu\!=\!0$ we have

	\begin{eqnarray}
	P_0&=&{1\over2}(E_{\bf p+q}\!+\!E_{\bf p})\ ,\nonumber\\
	&&\label{aad}\\
	r_0&=&2\gamma_5{\vec\gamma}\cdot({\bf p}\times{\bf q}).
	\nonumber\end{eqnarray}

\noindent For matrix elements between mean-field spinors, we have for
$\mu\!=\!0$:

	\begin{equation}\Gamma_0={1\over1+\tilde\tau^*}\Biggl[
	G_E(Q^2)\ {(E_{\bf p+q}^*\!+\!E_{\bf p}^*)\over2M^*}
	+ G_M(Q^2)\ {\gamma_5{\vec\gamma}\cdot({\bf p}\!\times
	\!{\bf q})\over2{M^*}^2}\Biggr].\label{aae}\end{equation}

\noindent In Model G, we use the on-shell forms of $G_E(Q^2)$ and
$G_M(Q^2)$, given in (\ref{aae}).  In Model F, we use $G_{E\sigma}^*$
and $G_{M\sigma}^*$, given in (3.11).  Expression (A5) makes clear the
origin of the functional forms in (3.6) and (3.7).


\vskip 0.3 true in
\noindent{\Large\bf Appendix B:\ \ \ Coefficients of Factored Moment Expansion}
\vskip 0.2 true in
\setcounter{equation}{0}
\seceqab

In this appendix, we give the coefficients $r_{i\sigma}$, for even powers
through ${\cal O}({\bf p}^4)$ in the moment expansion.  These are written
as they appear in Model G, i.e., in terms of the usual nucleon charge
($\epsilon_\sigma$) and magnetic moment ($\mu_\sigma$):

	\begin{equation}r_{0\sigma}({\bf q})=\epsilon_\sigma^2\
	{E_{\bf q}^*+M^*\over2E_{\bf q}^*},\label{aba}\end{equation}

	\begin{equation}r_{2\sigma}({\bf q})=
	\epsilon_\sigma^2\ \Biggl[{-4{E_{\bf q}^*}^5
	+5{E_{\bf q}^*}^4M^*+2{E_{\bf q}^*}^2{M^*}^3-3{M^*}^5
	\over12{E_{\bf q}^*}^5{M^*}^2}\Biggr]+\,\mu_\sigma^2\
	\Biggl[{{E_{\bf q}^*}-M^*\over 3{E_{\bf q}^*}{M^*}^2}\Biggr],
	\label{abb}\end{equation}

    \begin{eqnarray}r_{4\sigma}({\bf q})&=&{1\over240{E_{\bf q}^*}^9
    {M^*}^4}\ {{E_{\bf q}^*}-M^*\over{E_{\bf q}^*}+M^*}\
	\Biggl[\epsilon_\sigma^2\
	\Bigl[64{E_{\bf q}^*}^9+29{E_{\bf q}^*}^8M^*\nonumber\\
	&&\nonumber\\
	&&\qquad-6{E_{\bf q}^*}^7{M^*}^2+6{E_{\bf q}^*}^6{M^*}^3
	+18{E_{\bf q}^*}^5{M^*}^4+24{E_{\bf q}^*}^4{M^*}^5\nonumber\\
	&&\nonumber\\
	&&\qquad+30{E_{\bf q}^*}^3{M^*}^6-90{E_{\bf q}^*}^2{M^*}^7
	-210{E_{\bf q}^*}{M^*}^8\Bigr]\nonumber\\
	&&\label{abc}\\
	&&+\,\mu_\sigma^2\ \Bigl[-64{E_{\bf q}^*}^9-24{E_{\bf q}^*}^8M^*
	+16{E_{\bf q}^*}^7{M^*}^2\nonumber\\
	&&\nonumber\\
	&&\qquad-16{E_{\bf q}^*}^6{M^*}^3
	-48{E_{\bf q}^*}^5{M^*}^4-24{E_{\bf q}^*}^4{M^*}^5\Bigr].
	\Biggr]\nonumber\end{eqnarray}

\noindent These forms are also applicable to Model F, upon making the
replacements $\epsilon_\sigma\!\rightarrow\!\epsilon_\sigma^*({\bf q})$
and $\mu_\sigma\!\rightarrow\!\mu_\sigma^*({\bf q})$ using (3.14).


\vfill
\eject


\vskip 0.3 true in
\noindent{\Large\bf Figure Captions}
\vskip 0.2 true in
\noindent Fig.~1.  Coulomb sum $\Sigma({\bf q})$ for a uniform system
of Dirac protons, at two values of $M^*$.  Two different evaluations
of the lowest-order RCSR are shown as indicated.\hfil\break
\hfil\break
\noindent Fig.~2.  RCSR evaluated for a uniform system of nucleons with
anomalous magnetic moments, for $M^*/M\!=\!1$.  The Coulomb sum
$\Sigma({\bf q})$ in (4.1) and $S_A({\bf q})$ in (4.2) are shown
as indicated.\hfil\break
\hfil\break
\noindent Fig.~3.  Same as Fig.~2, but for $M^*/M\!<\!1$ in Model G.
Results are shown for (a) $M^*/M\!=\!0.718$ and (b) $M^*/M\!=\!0.556$.
\hfil\break\hfil\break
\noindent Fig.~4.  Same as Fig.~3, but in Model F.  Note that
$S_{II}({\bf q})$ is calculated in two ways: the ``factored'' (dashed)
and ``consistent'' (similar dot-dashed) moment expansions.\hfil\break
\hfil\break
\noindent Fig.~5.  Electric and magnetic recoil functions,
$r_E({\bf p},{\bf q})$ and $r_M({\bf p},{\bf q})$, in a convenient
separation.  Results are shown for (a) $M^*/M\!=\!1$ and (b)
$M^*/M\!=\!0.556$.  The functions $r_E({\bf 0},{\bf q})$ (solid),
$\delta r_E({\bf q})$ (dashed) and $\delta r_M({\bf q})$ (dot-dashed)
are shown; $r_M({\bf 0},{\bf q})\!=\!0$. (Note the scale changes for
$\delta r_E$ and $\delta r_M$.)\hfil\break
\hfil\break
\noindent Fig.~6.  Effective form factors (squared) in Model F, for
$M^*/M\!=\!0.556$.   Solid curves are obtained by averaging (3.11) with
(3.12) over the Fermi sphere, and include Fermi corrections to all orders.
Dashed curves are ${\epsilon_\sigma^*}^2({\bf q})$ and
${\mu_\sigma^*}^2({\bf q})$ of (3.14), as used in the ``factored''
moment expansion.
\hfil\break

\vfill
\eject

\end{document}